\documentclass[%
 preprint,
 amsmath,amssymb,
 aps,
]{revtex4-1}

\usepackage{graphicx}
\usepackage{dcolumn}
\usepackage{microtype}
\usepackage{setspace}

\begin{document}

\preprint{APS/123-QED}

\title{Quantification of a propagating spin-wave-packet created \\
by an ultrashort laser pulse in a thin film of magnetic metal}

\author{S. Iihama$^1$}
\email{iihama@mlab.apph.tohoku.ac.jp}
\affiliation{$^1$Department of Applied Physics, Graduate School of Engineering, Tohoku University, Japan \\
$^2$WPI Advanced Institute for Materials Research, Tohoku University, Japan}
 
\author{Y. Sasaki$^1$}
\affiliation{$^1$Department of Applied Physics, Graduate School of Engineering, Tohoku University, Japan \\
$^2$WPI Advanced Institute for Materials Research, Tohoku University, Japan}

\author{A. Sugihara$^2$}
\affiliation{$^1$Department of Applied Physics, Graduate School of Engineering, Tohoku University, Japan \\
$^2$WPI Advanced Institute for Materials Research, Tohoku University, Japan}

\author{A. Kamimaki$^1$}
\affiliation{$^1$Department of Applied Physics, Graduate School of Engineering, Tohoku University, Japan \\
$^2$WPI Advanced Institute for Materials Research, Tohoku University, Japan}

\author{Y. Ando$^1$}
\affiliation{$^1$Department of Applied Physics, Graduate School of Engineering, Tohoku University, Japan \\
$^2$WPI Advanced Institute for Materials Research, Tohoku University, Japan}

\author{S. Mizukami$^2$}
\email{mizukami@wpi-aimr.tohoku.ac.jp}
\affiliation{$^1$Department of Applied Physics, Graduate School of Engineering, Tohoku University, Japan \\
$^2$WPI Advanced Institute for Materials Research, Tohoku University, Japan}
 




\date{\today}

\begin{abstract}
Coherent spin-wave generation by focused ultrashort laser pulse irradiation was investigated for a permalloy thin film at micrometer scale using an all-optical space and time-resolved magneto-optical Kerr effect.
The spin-wave packet propagating perpendicular to magnetization direction was clearly observed, 
however that propagating parallel to the magnetization direction was not observed. 
The propagation length, group velocity, center frequency, and packet-width of the observed spin-wave packet 
were evaluated and quantitatively explained in terms of the propagation of a magneto-static spin-wave driven by ultrafast change of an out-of-plane demagnetization field induced by the focused-pulse laser.
\end{abstract}

\pacs{Valid PACS appear here}
\maketitle



Information processing by spin-waves such as spin-wave logic gates or magnon transistors \cite{Kostylev, Schneider, Sato, Jamali,Chumak},
has attracted much attention for the realization of future devices with low power consumption. 
In previous studies, the propagating spin-waves have been excited by a local radiofrequency (rf) magnetic field generated by an  rf electric current fed into an antenna \cite{Covington, Wu, Tamaru, Sekiguchi, Yu, Birt, Sebastian, Schneider2008} or by a dc electric current via a spin-transfer-torque \cite{Demidov, Madami}.
Among them, manipulation of spin-waves propagating in thin films of magnetic metals are of interest \cite{Covington, Tamaru, Sekiguchi, Yu, Birt, Sebastian},
because magnetic metal thin films may be well-suited for integration into silicon nano-technology \cite{Serga}.
However, the propagation length of spin-waves in thin films of magnetic metals is very short,
so that the precise characterization of the spin-wave propagation in the near sub-micrometer region is a technological challenge, for which
it may be valuable to explore the various observational methods.

Recently, it has been demonstrated that a propagating spin-wave can be generated by a focused-ulrashort laser pulse in an insulating ferrimagnet \cite{Satoh}. 
This finding is the beginning of {\it opto-magnonics}, which opens new avenues for investigating propagating spin-waves using laser light. 
Inspired by this experiment, Au {\it et al.} also attempted to observe the propagating spin-wave created in metallic magnetic thin films by a laser pulse and discussed the nature of spin-wave excitation.
However, their discussion was still qualitative because the strong heating and acoustic wave induced by laser pulse irradiation overlapped the spin-wave propagation, making quantitative analysis difficult to perform \cite{Au}.
Subsequently Yun {\it et al.} also performed a similar experiment and evaluated the spin-wave propagation length in CoFeB thin films,
but also did not present a quantitative discussion for the mechanism behind the generation of spin-waves by a pulse laser \cite{Yun}.
Thus, this mechanism has yet to be clarified.
Generally, there are several discussions on the laser-induced spin dynamics, even in insulating magnets in which the inverse Faraday effect is considered to be the main origin of spin-wave generation \cite{Kimel}.
In the case of a metallic magnet, strong light absorption and ultrafast demagnetization occur over a very short-time scale when the laser pulse is incident upon it \cite{Beaurepaire, Koopmans}; 
this is an unique property in metallic magnets, from which it follows that  the scenario of spin-wave generation by a pulse laser may differ between metals and insulators.

In this letter, we report a clear observation of the propagation of a spin-wave packet created by an ultrashort laser pulse in a permalloy thin film by means of the space- and time-resolved magneto-optical Kerr effect (STR-MOKE).
The observed propagation of a spin-wave generated by a laser pulse is {\it quantitatively} explained by a theoretical model taking account of the ultrafast change of the out-of-plane demagnetization field induced by the laser pulse with a Gaussian distribution in space.

The STR-MOKE measurements were performed by a standard, all-optical pump-probe set-up with the two-color method [Fig. 1(a)]. 
The source of the pulse was a Ti: Sapphire laser with a wavelength, pulse width, and pulse repetition rate of $\sim$800 nm, $\sim$120 fs, and $\sim$80 MHz, respectively.
A BaB$_2$O$_4$ (BBO) crystal was used for the frequency-doubling of the pump laser. 
Then, the pump laser amplitude was modulated using a mechanical chopper with a modulation frequency at 370 Hz.
Both the probe and the pump laser with linear-polarization were focused onto a sample by the objective lens with $NA=0.65$ via a scanning mirror by which 
the spot position of the probe laser beam on the sample can be changed by changing the angle of the mirror.
The spot radius of the probe laser beam ($\sigma _{\rm probe}$), {\it i.e.,} the half-width at which the light intensity becomes $e^{-0.5}$, 
was evaluated to be 0.51 ${\rm \mu m}$ using the knife-edge method \cite{Araujo}. 
The intensity of the pump laser beam at the focused area approximated to a Gaussian function with radius ($\sigma _{\rm pump}$) was evaluated to be 0.50 $\mu$m \cite{Sup}; the fluence of the pump laser pulse was estimated to be several mJ/cm$^2$.
The spot position of the probe laser beam on the sample was scanned along the direction perpendicular ($x$-axis) or parallel ($y$-axis) to the in-plane component of the magnetization $\mbox{\boldmath $M$}$ [Fig. 1(b)].
The pump laser pulse-induced change in the Kerr rotation angle of the reflected probe laser pulse ($\Delta \varphi _{\rm K}$) was detected by a balanced photo-diode detector and a lock-in amplifier,
and the output voltage was then recorded as a function of the pump-probe delay time and distance.
During the measurements, a magnetic field $\mbox{\boldmath $H$}_{\rm ext}$ of 3 kOe  with a field angle of $\sim$5$^{\circ }$ from the normal of the sample plane ($z$-axis) was externally applied by a permanent magnet to the sample.
The sample consists of a 20-nm-thick permalloy (Ni$_{80}$Fe$_{20}$) film deposited onto a Si/SiO$_2$ substrate by ultra-high  vacuum magnetron sputtering.

%
\begin{figure}
\begin{center}
\includegraphics[width=8.2cm,keepaspectratio,clip]{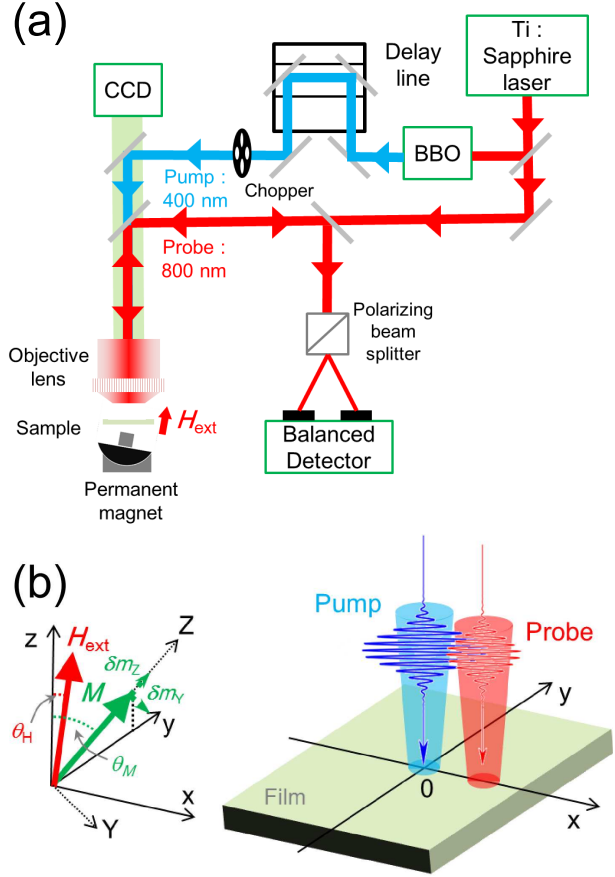}
\caption{Schematic illustration of (a) the optical set-up for the space- and time-resolved magneto-optical Kerr effect 
and (b) the coordinate system. 
The probe beam spot was scanned along the direction perpendicular ($x$ direction) or parallel ($y$ direction) to the in-plane component of 
the magnetization unit vector, $\mbox{\boldmath $M$}$.
Both  $\mbox{\boldmath $M$}$ and the external magnetic field vector $\mbox{\boldmath $H$}_{\rm ext}$ are in the $y-z$ plane.
The directions of $\mbox{\boldmath $H$}_{\rm ext}$ and $\mbox{\boldmath $M$}$ are defined by the angles $\theta _{\rm H}$ and $\theta _{\rm M}$ with respect to a film normal, respectively.
$\delta m_{\rm Z}$ and $\delta m_{\rm Y}$ correspond to small changes in the parallel and perpendicular components of $\mbox{\boldmath $M$}$, respectively, in an $X-Y-Z$ coordinate system that is defined by the rotation of the $x-y-z$ coordinate system about the $x$-axis.  
}
\label{f1}
\end{center}
\end{figure}  
%
%
\begin{figure}[h]
\begin{center}
\includegraphics[width=8cm,keepaspectratio,clip]{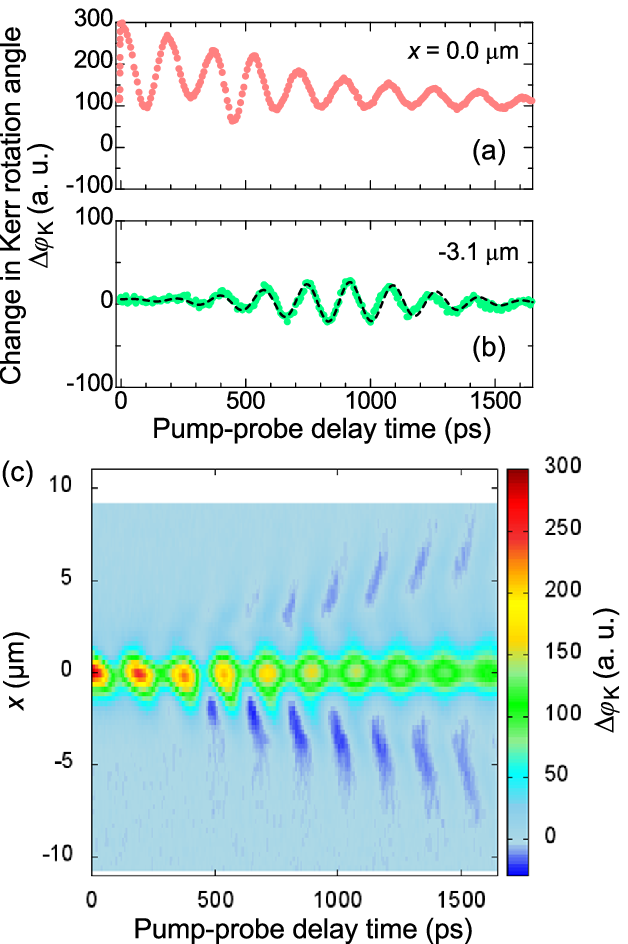}
\end{center}
\caption{Change in the Kerr rotation angle $\Delta \varphi _{\rm K}$ as a function of time at $x$ = 0 $\mu $m (pump-probe overlap) (a) and $x$ = -3.1 $\mu $m (b). Broken curves are fitted to results using a Gaussian wave-packet. (c) The value of $\Delta \varphi _{\rm K}$ is mapped as a function of time and probe position $x$.}%
\label{f2}
\end{figure}

Figures 2(a) and 2(b) show $\Delta \varphi _{\rm K}$ as a function of time with $x=$ 0 and -3.4 $\mu$m, respectively.
$\Delta \varphi _{\rm K}$ at $x=0$ $\mu$m exhibits a damped-sinusoidal oscillation superposed with an exponentially-damped background. 
The damped-oscillation and background are, respectively, attributed to the damped-magnetization precession and the recovery of magnetization magnitude from ultrafast demagnetization, and are both induced by the pump laser over a short time scale \cite{Kampen2002}.
On the other hand, the shape of the $\Delta \varphi _{\rm K}$ observed at $x=$ -3.4 ${\rm \mu m}$ becomes a wave-packet form [Fig. 2(b)]. 
The pump beam is not incident at this area, 
so this spatio-temporal non-local pump-probe response evidently demonstrates the spin-wave packet propagation generated by the pump laser pulse.
Spatio-temporal evolution of a spin-wave is more visible in the $\Delta \varphi _{\rm K}$ mapped in space-time [Fig. 2(c)].
The large $\Delta \varphi _{\rm K}$ around $x = 0$ $\mu$m corresponds to the change of magnetization magnitude as mentioned above.
The strip-like pattern of $\Delta \varphi _{\rm K}$ around $x = 0$ $\mu$m corresponds to the emission and time-evolution of the wave.
The center of mass of the wave-packet symmetrically moves up to about 5 $\mu$m within 1.5 ns from the pump laser irradiated area, 
indicating that the group velocity of the spin-wave packet is about 3-4 km/s.
On the other hand, no propagating spin-wave was clearly observed when the scanning direction of the probe laser beam was parallel to the $y$-axis (not shown here).
\begin{figure}
\begin{center}
\includegraphics[width=8cm,keepaspectratio,clip]{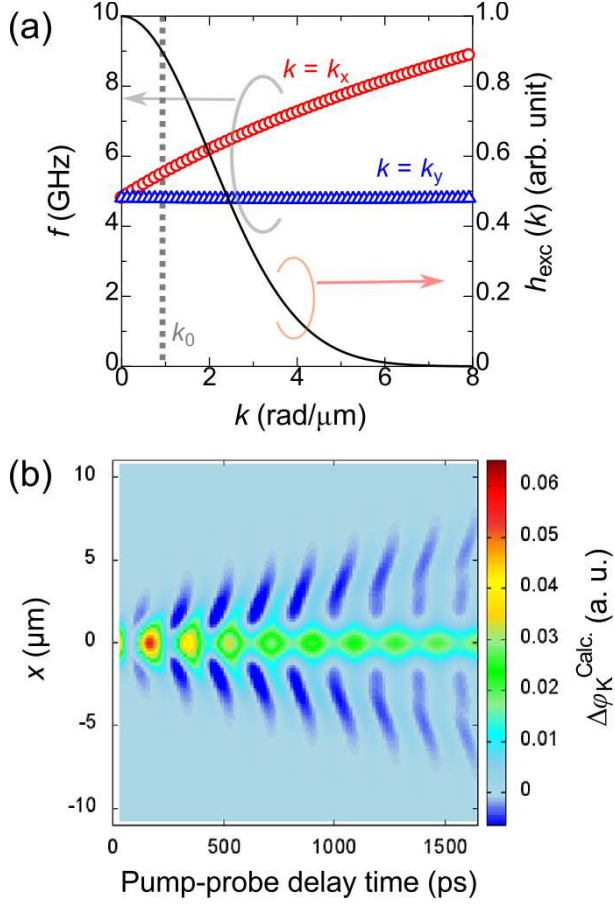}
\end{center}
\caption{(a) The spin-wave frequency $f$ as a function of the wave-number $k$ calculated for the 20-nm-thick permalloy film under an oblique magnetic field for the $k_{\rm x}$ (red open circles) and $k_{\rm y}$ directions (blue open triangles), respectively.
An effective excitation magnetic field in $k$-space $h_{\rm exc}(k)$ was also plotted (solid curve),
which is generated by the ultrafast demagnetization induced by a focused pump laser pulse. 
(b) The simulated value of $\Delta \varphi _{\rm K}$ is mapped as a function of time and probe position $x$, and corresponds to that in Fig. 2(c).
}
\label{f3}
\end{figure} 

The-above-mentioned value of the group velocity and its anisotropy are qualitatively consistent with the nature of long-wavelength spin-waves  in magnetic films, {\it i.e.,} magnetostatic spin-waves.
The magnetostatic spin-wave is governed not by a short-range exchange interaction, but by a long-range magnetic dipolar interaction,
so that the two anisotorpic modes depending on their propagation directions are known in in-plane magnetized films, 
{\it i.e.,} the magnetostatic surface spin-wave mode (MSSW) and the magnetostatic backward volume wave mode (MSBVW)with the wave vector perpendicular and parallel to the magnetization direction, respectively.
The group velocity of MSSW is generally much larger than that of MSBVW in the case of thin films; 
thus, the observed spin-wave packet may be attributed to that of the MSSW created by a laser pulse.
In order to confirm this speculation, quantitative analysis is necessary with the magnetostatic spin-wave dispersion relation and the details of the laser-induced spin-wave generation mechanism taken into account.
This is because the nature of the spin-wave observed should depend on its average frequency and wave number, 
which are determined by the dispersion relation and the wave number-dependent stimulus by the pulse laser.

The coherent spin-wave generated by a pulse laser may be described by linear response theory based on the linearized Landau-Lifshitz-Gilbert equation taking account of the magnetostatic interaction with a thin film limit (see the details of the calculation in the Supplementary Materials\cite{Sup}) :
\begin{equation}
{\small
\delta m_i (\mbox{\boldmath $r$},t) = \iint  \tilde{\chi }_{ij} (\mbox{\boldmath $r$}-\mbox{\boldmath $r$}^{\prime },t-t^{\prime })h_{\rm exc,  \it j} (\mbox{\boldmath $r$}^{\prime },t^{\prime }) d^2r^{\prime }dt.
} 
\end{equation}
Here, $\delta m_i$, $\tilde{\chi }_{ij}$, and $h_{\rm exc,  \it j}$ ($i, j =X, Y$), as the coordinates defined in Fig. 2(b), are the small deviation  from $\mbox{\boldmath $M$}$,
the rf magnetic susceptibility tensor including spin-wave dispersion relation, and the rf external magnetic field in space $\mbox{\boldmath $r$}=(x,y)$ and time $t$, respectively.
The spin-wave dispersion relation in the case of arbitrary magnetization angle was calculated by using the formula described in Ref. \cite{McMichael, Landeros}.
Here, we assume that $\mbox{\boldmath $h$}_{\rm exc}$ is attributed to the temporal change of an out-of-plane demagnetization field induced by ultrafast demagnetization in the focused area, which can be expressed as
\begin{eqnarray}
&& h_{\rm exc, \it Y}(\mbox{\boldmath $r$},t)= -4 \pi M_{\rm s} \delta m_Z (\mbox{\boldmath $r$},t) \sin 2\theta _{\rm M},  \\
&& \delta m_Z(\mbox{\boldmath $r$},t) =  \left[ \frac{\Delta M_{\rm s}^0}{M_{\rm s}} \exp \left(-\frac{t}{\tau _{\rm rec}}\right) \Theta (t)+C_0\right] \nonumber \\
&&\hspace{4cm}\times \exp \left(-\frac{r^2}{2\sigma _{\rm pump}^2}\right). 
\end{eqnarray}
Here, we approximate the saturation magnetization $M_{\rm s}$ as being instantaneously reduced when the pump pulse laser shines and recovering exponentially with the recovery time $\tau _{\rm rec}$.
$\Delta M_{\rm s}^0$, $\Theta (t)$, and $C_0$ represent the magnitude of ultrafast demagnetization, the Heaviside step function, and the reduction of saturation magnetization due to the time-averaged temperature rising.
These equations are physically interpreted as follows.
The quantities relevant to propagation of the spin-wave are determined by the dispersion relation of the spin-wave and the excitation magnetic field in the wave-number $\mbox{\boldmath $k$}$ space, which are computed from $\tilde{\chi}$ in Eq. (1) and the Fourier transform of Eq. (2), respectively [Fig. 3(a)].
The respective MSSW and MSBVW modes show approximately linear and flat dispersion
in the region of  $k < 6$ ${\rm rad/\mu m}$, where the Fourier transform of $h_{\rm exc, \it Y}$ has finite strength.
Spin-waves with the $\mbox{\boldmath $k$}$ vector in this region are excited in-phase at $t=0$ 
and then form the wave-packet propagating away from the origin.
The excitation source, {i.e.,} $h_{\rm exc, \it Y}$, is symmetric in space but the spin-wave dispersion is highly anisotropic,
so that the spin-wave directionally propagates, as mentioned above. 

The observed $\Delta \varphi _{\rm K}$ value may be proportional to the normal component $\delta m_z (r, t)$, 
which can be computed by taking account of the finite radius of the probe beam spot:
\begin{eqnarray}
&& \Delta \varphi_{\rm K}(\mbox{\boldmath $r$},t) \propto \iint \delta m_z (\mbox{\boldmath $r$}-\mbox{\boldmath $r$}^{\prime },t) G_{\rm probe}(\mbox{\boldmath $r$}^{\prime })d^2r^{\prime }, \label{eq:obs} \\
&& \delta m_z (\mbox{\boldmath $r$},t)=\delta m_Y (\mbox{\boldmath $r$},t)\sin \theta _{\rm M} + \delta m_Z (\mbox{\boldmath $r$},t) \cos \theta _{\rm M}, \label{eq:dmz}
\end{eqnarray}
where $G_{\rm probe}$ in Eq. (\ref{eq:obs}) denotes Gaussian intensity distribution of probe beam, first and second terms in Eq. (\ref{eq:dmz}) correspond to the oscillating component and the reduction of magnetization, respectively. 
%
The calculated $x$-$t$ mapping of $\Delta \varphi_{\rm K}$ is also shown in Fig. 3(b).
The data agrees well with the experimental data shown in Fig. 2(c).
The calculated data of the $y$-$t$ mapping of $\Delta \varphi_{\rm K}$ did not show a propagating spin-wave pattern (not shown here),
which was also consistent with experimental results.

\begin{figure}
\begin{center}
\includegraphics[width=8.5cm,keepaspectratio,clip]{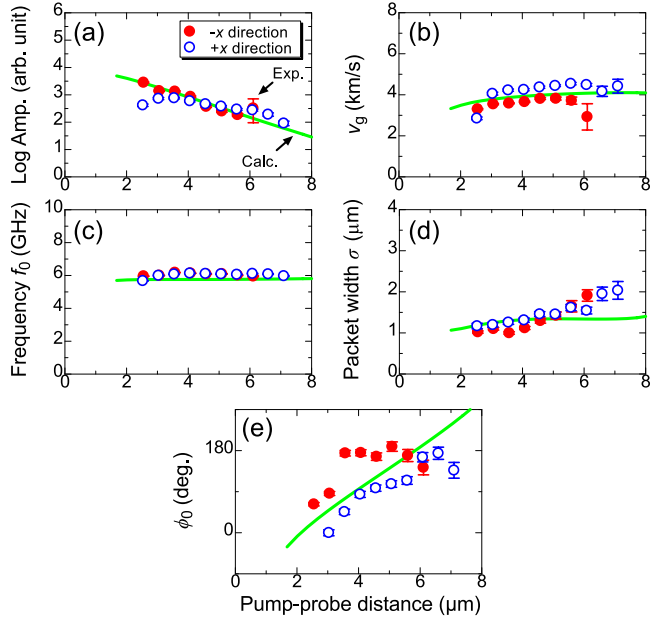}
\caption{Logarithmic amplitude (a), group velocity $v_{\rm g}$ (b), frequency $f_0$ (c), packet width $\sigma $ (d), and initial phase $\phi _0$ (e) of the spin-wave obtained from the fitting as a function of the distance between the pump and probe positions. Solid and open symbols represent the values evaluated from the -$x$ and +$x$ directions, respectively. The solid curve indicates the calculated result.}
\label{f4}
\end{center}
\end{figure}


In order to quantitatively compare the calculated and experimental data,
the theoretical and experimental values of $\Delta \varphi_{\rm K}$ corresponding to the spin-wave packet were analyzed 
using the fitting of the Gaussian wave-packet:
\begin{eqnarray}
\Delta \varphi_{\rm K}(x,t)&=&A \exp \left[ \frac{\left( v_{\rm g}t-x\right) ^2}{2\sigma ^2}\right] \sin \left( 2\pi f_0 t -\phi _0 \right). \label{eq:fit}
\end{eqnarray}
Here $A$, $v_{\rm g}$, and $\sigma $ are the amplitude, group velocity, and packet width, respectively, of a Gaussian envelope,
and $f_0$ and $\phi _0$ are the center frequency and phase of the spin-wave, respectively, for a sinusoidal function.
Generally, a spin-wave packet generated by a pulse magnetic field forms a shape slightly deviated from the Gaussian wave-packet \cite{Wu},
whereas the measured wave-packets are approximately fitted by Eq. (\ref{eq:fit}), as shown in Fig. 2(a) with a broken curve. 
The full set of quantities obtained from the experimental and theoretical data are shown in Fig. 4 
with circles and solid curves, respectively.
The experimental values agree well with the theoretical data within acceptable errors.

The propagation length, $\lambda $, of the spin-wave was evaluated to be 3.5 $\pm $ 0.1 ${\rm \mu m}$ from the data in Fig. 4(a),
which is relatively shorter than the values reported in a permalloy thin films.
The $\lambda $ value may be simply connected to $v_{\rm g}$ and the relaxation time of the spin-wave, $\tau $, by the relation $\lambda = v_{\rm g} \tau $.
When the Gilbert damping is the main loss mechanism, 
the $\tau $ value is roughly estimated by the relation $\tau  \sim 1/ 2 \pi \alpha f_0 P$, where $\alpha $ and $P$ are the Gilbert damping constant and the precession ellipticity factor, respectively.
The $v_{\rm g}$ and $f_0$ values evaluated from the experiment are roughly independent of $x$ and are about 4 km/s and 5.5 $\sim $ 5.9 GHz, respectively [Figs. 4(b) and 4(c)]. The $P$ value was calculated to be $\sim $ 2.6.
These values yield $\tau \sim$ 1.6 ns and $\lambda = 6.4$ $\mu$m with the experimental $\alpha$ of 0.0071
 being approximately consistent with the-above-mentioned $\lambda$ values. 
The packet width, $\sigma$, is $\sim$1 $\mu$m [Figs. 4(d)], 
which is comparable to the width of the effective excitation field in $k$-space [Fig. 3(a)].
The center wave-number, $k_0$, for the generated spin-wave can be roughly 
evaluated to be 0.8 $\sim $ 1.0 rad/$\mu $m by the slope, {\it i.e.,} $\phi_0 = k_0 x$, from the data in Fig. 4(e). 
The values of $f_0$ and $v_{\rm g}$ ($=2\pi df/dk$) at $k_0$ for the spin-wave dispersion in the film, as shown in Fig. 3(a) by the vertical broken line, 
also accord with the values shown in Figs. 4(b) and 4(c), respectively. 

The above discussion allows us to conclude that the primary origin of the coherent spin-wave generated by the focused pulse laser in permalloy films
is the local modulation of the out-of-plane demagnetization field, {\it i.e.,} Eq. (2).
However, Eq. (2) is a rather naive approximation of the effective magnetic field.
Another contribution resulting from the local reduction of magnetization is the effective magnetic field stemming from magnetic charges along the rim of the pump laser-focused spot.
This has a spatial symmetry different from that expressed by Eq. (2) and results in an antisymmetric spin-wave emission with respect to the magnetization direction, 
as suggested by the theoretical simulation by Au {\it et al.} \cite{Au}.
The phase of the spin-wave observed in this study is close to symmetric [Fig. 2(b)],
so that the effect of such contributions may be ruled out by the result observed here.
The phenomenon should be investigated in more detail by means of a two-dimensional imaging of the spin-wave propagation, with the simulation taking a more correct derivation of Eq.(2) into account using the Poisson equation, or more generally, Maxwell's equations with the metal's finite conductivity,
which will be addressed as a next step. 

In conclusion, we have investigated an ultrashort laser pulse-induced propagating spin-wave in a 20-nm-thick permalloy film. 
The propagating spin-wave packet was clearly observed in the case where the propagation direction was perpendicular to the magnetization direction. 
Various quantities that characterize the propagating spin-wave were experimentally evaluated, and those were quantitatively well-explained by the theoretical model
in which the propagating spin-wave was excited by the out-of-plane demagnetization field with a Gaussian intensity distribution that was induced by the focused pulse laser.

This  work  was  partially  supported  by Grants-in-Aid for Scientific Research (Nano Spin Conversion Science, No. 26103004)
and the WPI-AIMR fusion research project.
S. I. and S. M. thank to the Neo-Arc corp., Japan and Y. Kondo for their assistances for development of experimental set-up. 
They also thank to S. Tamaru, T. Satoh, Y. Hashimoto, and T. Miyazaki for valuable discussions. 
S. I. and Y. S. thank to Grant-in-Aid for JSPS Fellow (No. 26-4778) and to the GP Spin program, respectively.

\end{document}